\documentclass[letterpaper, 10 pt, conference]{ieeeconf}
\IEEEoverridecommandlockouts
\usepackage{booktabs}
\usepackage{multirow}
\usepackage{caption}
\usepackage[dvipsnames]{xcolor}
\usepackage{verbatim}
\usepackage{graphicx}
\usepackage{amsmath}
\usepackage{hyperref}
\usepackage{caption}
\usepackage{subcaption}
\usepackage{amsmath,amssymb,amsfonts,mathrsfs}
\usepackage{etoolbox}
\usepackage[normalem]{ulem}
\DeclareMathOperator*{\argmin}{\arg\!\min}
\graphicspath{{./figures/}}
\title{\LARGE \bf
A GPU Implementation of a Look-Ahead Optimal Controller for Eco-Driving Based on Dynamic Programming
}
\author{Zhaoxuan Zhu$^{*}$, Shobhit Gupta, Nicola Pivaro, Shreshta Rajakumar Deshpande and Marcello Canova%
\thanks{$^{*}$Corresponding author. Email: {\tt\small zhu.1083@osu.edu}}%
\thanks{The authors are with the Center for Automotive Research, The Ohio State University, Columbus, OH 43212, USA }%
\thanks{This version has been accepted for publication in Proc. European Control Conference (ECC), 2021. Personal use of this material is permitted. Permission from EUCA must be obtained for all other uses, in any current or future media, including reprinting/republishing this material for advertising or promotional purposes, creating new collective works, for resale or redistribution to servers or lists, or reuse of any copyrighted component of this work in other works.}
}
\begin{document}
\maketitle
\thispagestyle{empty}
\pagestyle{empty}
\begin{abstract}
Predictive energy management of Connected and Automated Vehicles (CAVs), in particular those with multiple power sources, has the potential to significantly improve energy savings in real-world driving conditions. In particular, the eco-driving problem seeks to design optimal speed and power usage profiles based upon available information from connectivity and advanced mapping features to minimize the fuel consumption between two designated locations. 

In this work, the eco-driving problem is formulated as a three-state receding horizon optimal control problem and solved via Dynamic Programming (DP). The optimal solution, in terms of vehicle speed and battery State of Charge (SoC) trajectories, allows a connected and automated hybrid electric vehicle to intelligently pass the signalized intersections and minimize fuel consumption over a prescribed route. To enable real-time implementation, a parallel architecture of DP is proposed for an NVIDIA GPU with CUDA programming. Simulation results indicate that the proposed optimal controller delivers more than 15\% fuel economy benefits compared to a baseline control strategy and that the solver time can be reduced by more than 90\% by the parallel implementation when compared to a serial implementation.
\end{abstract}

\section{INTRODUCTION}
Connected and Automated Vehicles (CAVs) have the potential to increase safety, driving comfort as well as fuel economy, by exploiting features such as advanced mapping, GPS location, and information available via vehicle-to-vehicle (V2V) and vehicle-to-infrastructure (V2I) communication \cite{guanetti2018control, gupta2020estimation}. Meanwhile, Hybrid Electric Vehicles (HEVs) increase the overall powertrain efficiency by including battery pack(s) and electric motor(s) as alternative energy storage and power generation devices \cite{guzzella2007vehicle}. Combining the two technologies could further improve the fuel economy, however, poses a greater challenge from the planning\&control perspective. 

In this context, finding the optimal powertrain control strategy that minimizes the total fuel consumption and the travel time between origin and destination is known in literature as the eco-driving problem \cite{sciarretta2015optimal}. The contributions made in this field distinguish among two aspects, namely powertrain configurations and traffic scenarios. Regarding powertrain configuration, the difference is in whether the powertrain is equipped with a single power source \cite{ozatay2014cloud, jin2016power, han2019fundamentals, sun2020optimal} or features a hybrid architecture \cite{guo2016optimal, olin2019reducing, bae2019real}. The latter requires a more complex control algorithm, as the battery State-of-Charge (SoC) needs to be regulated and utilized efficiently. 

The difference in traffic scenarios lies in whether the controller is capable of processing the real-time Signal Phase and Timing (SPaT) information at signalized intersections. Ozatay et al. \cite{ozatay2014cloud} proposed a framework providing advisory speed profile using online optimization conducted on a cloud-based server without considering the real-time traffic light variability. Olin et al. \cite{olin2019reducing} used Dynamic Programming (DP) to solve the eco-driving problem. As traffic lights are not explicitly considered in these studies, the control logic requires assistance from other decision-making agents, such as human drivers or adaptive cruise control (ACC) systems. Other studies have explicitly modeled and considered SPaT, for example Jin et al. \cite{jin2016power} formulated the problem as a Mixed Integer Linear Programming (MILP) for a conventional vehicle. Asadi et al. \cite{asadi2011predictive} used traffic simulation models and proposed to solve the problem considering probabilistic SPaT with DP. Sun et al. \cite{sun2020optimal} formulated the eco-driving problem as a distributionally robust stochastic optimization problem with collected real-world data and solved it with DP. Guo et al. \cite{guo2016optimal} proposed a hierarchical control framework with a hybrid vehicle. Bae \cite{bae2019real} extended the work in \cite{sun2020optimal} by including a heuristic HEV supervisory controller. 

Many of the aforementioned studies used DP \cite{bertsekas2005dynamic} as the technique to either solve the optimization within the receding horizon of a MPC or the entire problem. Unfortunately, the existing use of DP either focus on enabling the vehicle with single power source to pass the signalized intersections \cite{asadi2011predictive, sun2020optimal} or on optimally splitting the power demand between engine and electrical motor \cite{olin2019reducing} without considering SPaT information. In \cite{sun2020optimal}, two states, namely vehicle speed and travel time, are included in their DP formulation, whereas in \cite{olin2019reducing}, velocity and SoC are considered as the DP states. Since the number of DP states is subject to curse of dimensionality \cite{bertsekas2005dynamic}, including one extra state often results in intractable computational requirements, preventing from online implementation in rapid prototyping control systems. 

In this study, the DP formulation proposed in \cite{sun2020optimal} and \cite{olin2019reducing} are merged to develop an eco-driving control strategy for HEVs that is also able to optimally cross signalized intersections, given real-time SPaT information. To mitigate the computational burden from the additional state, an efficient parallel implementation is developed for a NVIDIA GPU with CUDA programming.


\section{Model Development and Validation} \label{sec: model development}
In this work, a forward-looking dynamic powertrain model is utilized for performance and fuel economy prediction over real-world routes \cite{olin2019reducing}. The model considers a P0 mild-HEV with a 48V Belted Starter Generator (BSG) performing torque assist, regenerative braking and start-stop functions.  

\begin{figure}[t!] 
	\centering
	\vspace*{0.15cm}\includegraphics[width=\columnwidth]{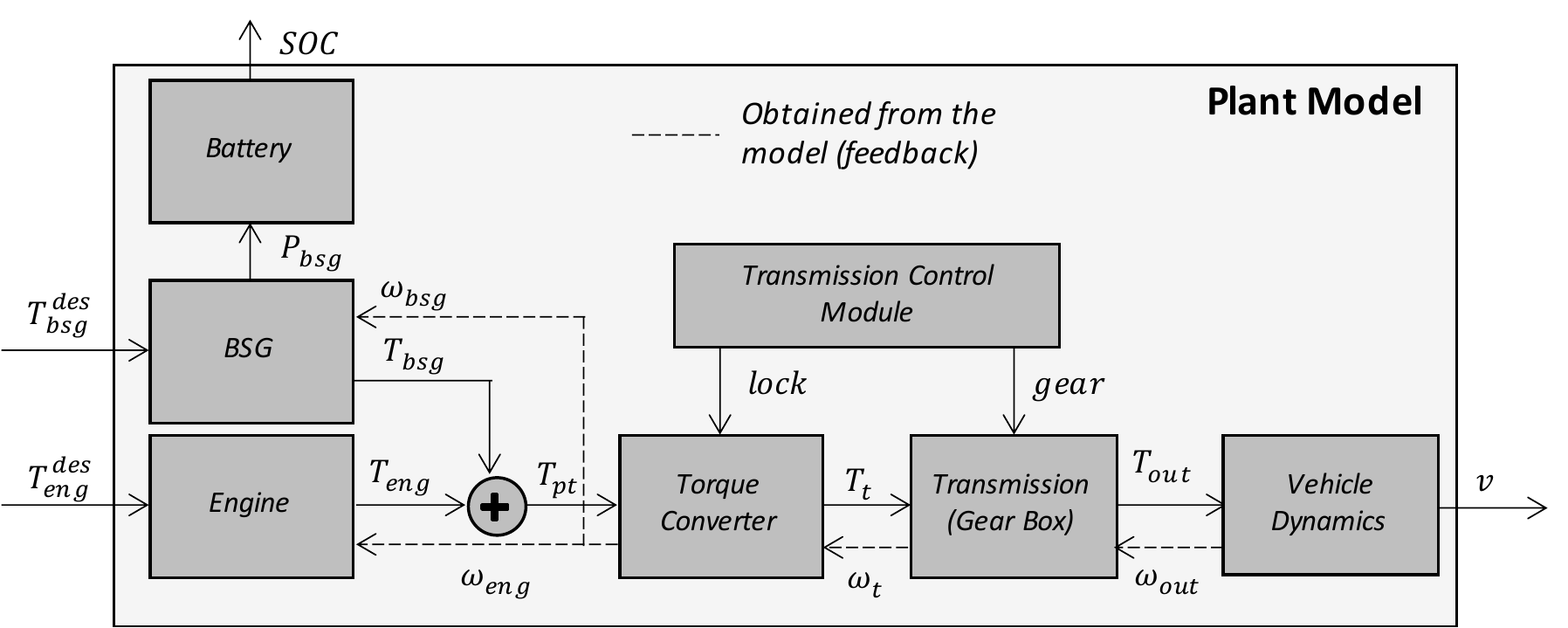}
	\caption{Block Diagram of 48V P0 Mild-Hybrid Drivetrain.}
	\label{fig_plant_model}
\end{figure}

As shown in Fig. \ref{fig_plant_model}, the inputs to the vehicle dynamics and powertrain (VD\&PT) model are obtained from a simplified model of the Engine Control Module (ECM), which contains the essential functions to convert the driver’s input (pedal position) to torque commands. The outputs of the ECM, the desired BSG torque ($T^{\mathrm{des}}_{\mathrm{bsg}}$) and desired engine torque ($T^{\mathrm{des}}_{\mathrm{eng}}$) are obtained from a production torque split strategy, which is used as the baseline for fuel economy evaluation.

The vehicle simulator contains low-frequency dynamic models of the powertrain and longitudinal vehicle dynamics. The battery is modeled as a zero-th order equivalent circuit, from which the battery State-of-Charge (SoC) is calculated; the vehicle longitudinal dynamics are described by the road load equation \cite{guzzella2007vehicle}. Low-frequency, quasi-static models are developed for the engine (fuel and friction maps), BSG, torque converter and transmission (efficiency maps).

\begin{figure}[t!]
	\centering
	\vspace*{0.15cm}\includegraphics[width=\columnwidth]{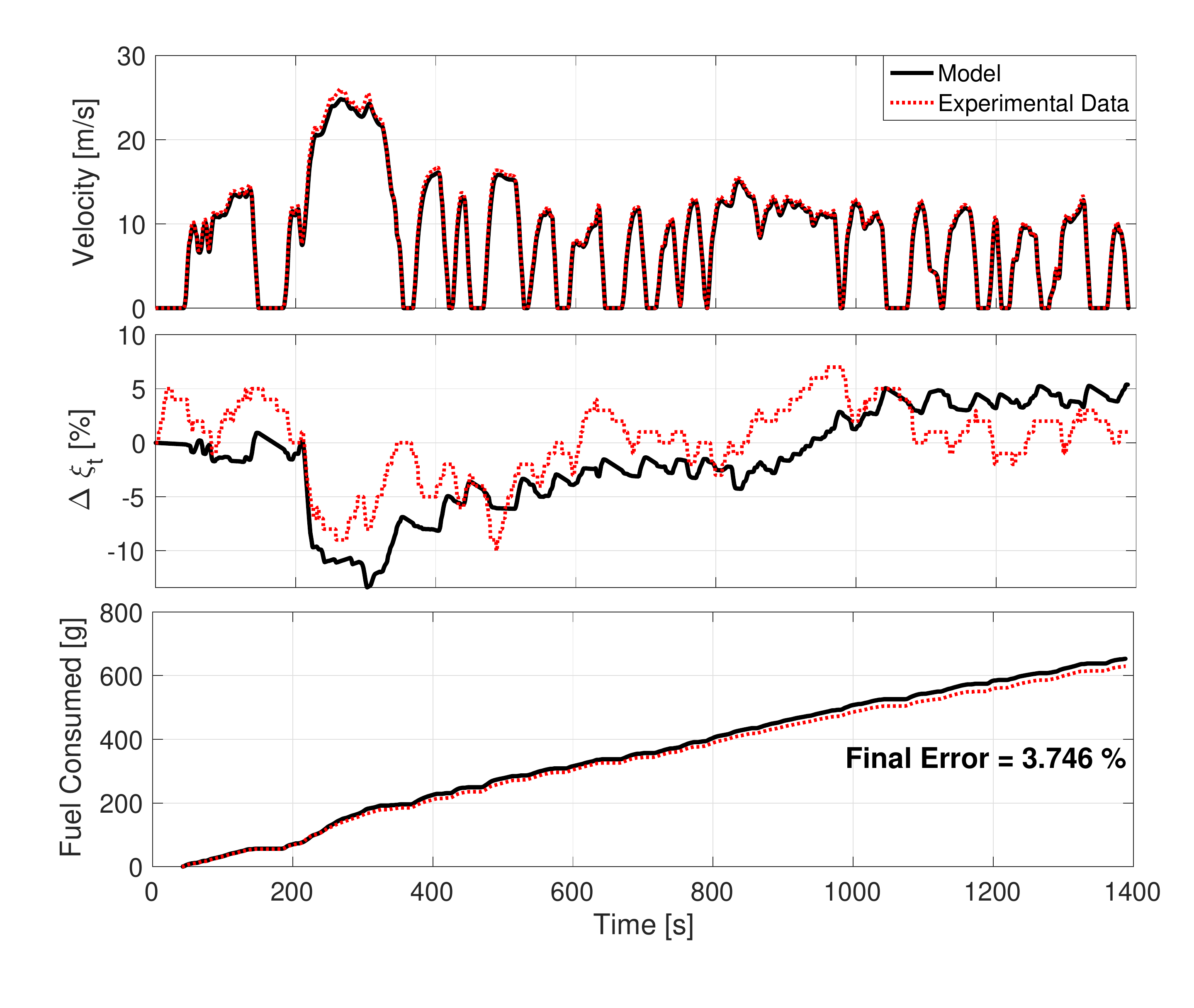}
	\caption{Validation of Vehicle Velocity, SoC and Fuel Consumed over FTP-75.}
	\label{fig_veh_vel_soc_fuel_validation_FTP}
\end{figure}

The vehicle model was validated based on chassis dynamometer data. Fig. \ref{fig_veh_vel_soc_fuel_validation_FTP} shows a verification over the drive cycle FTP-75, where the vehicle velocity, battery SoC ($\xi$) and fuel consumption are compared against test data. Small mismatches in the battery SoC profiles are attributed to simplification of the 12V electrical system, where the auxiliary loads are modeled using a constant current bias. The fuel consumption over FTP-75 is well predicted, with cumulative error less than 4\%. 

\section{Problem Formulation and Solution Methods} \label{sec: problem formulation}
The objective of the eco-driving problem, formulated in the spatial domain, is to minimize the fuel consumed by the vehicle over an entire route consisting of N steps:
\begin{equation}
    \min_{T_{\mathrm{eng},s}, T_{\mathrm{bsg},s}} \sum_{s=1}^{s=N}{\left(\gamma \cdot \dot{m}_{\mathrm{f},s} +(1-\gamma)\right)\cdot \Delta t_s},
\end{equation}
where $s$ is the discrete index for the vehicle longitudinal position, $\gamma$ is the weighing factor between the fuel consumption and the travel time, and $\dot{m}_{\text{f},s}$ is the instantaneous fuel consumption. $\Delta t_s$ is the travel time per step computed as follows:
\begin{equation}
    \Delta t_s=\frac{\Delta d}{\bar{v}_{s}},
\end{equation}
where $\Delta d$ is the distance step, i.e. $\Delta d=d_{s+1}-d_s$, calculated from $d_s$ the distance traveled along the route, and $\bar{v}_{s}=\frac{v_{s}+v_{s+1}}{2}$ is the average velocity. The state and action space are subject to the constraints as described in \cite{olin2019reducing}.

A primary benefit of the spatial formulation \cite{saerens2012optimal} is that it inherently aids in the incorporation of distance-based route features, such as speed limits, grade, traffic light and stop sign locations. However, the spatial-domain formulation would make it difficult to incorporate time-based information such as SPaT received from V2I communication. This requires one to augment the state space with time as an additional state, which will be elaborated later in this section. 

\begin{figure}[t!]
	\centering
	\vspace*{0.15cm}\includegraphics[width=\columnwidth]{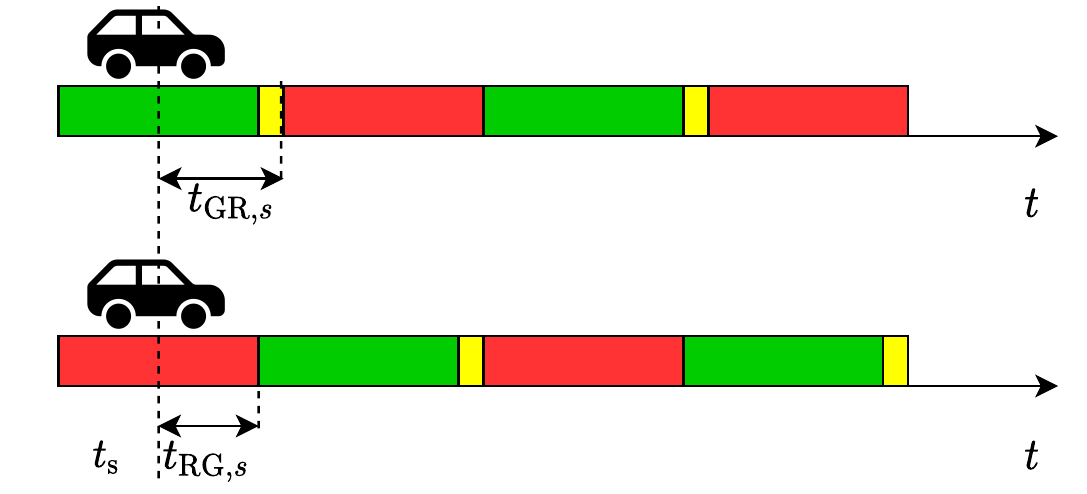}
	\caption{Possible Cases of Traffic Light Phase Encountered by the Vehicle.}
	\label{fig_Traffic_Light_Phase}
\end{figure}

Consider a case where a CAV is approaching a signalized intersection and would receive a traffic light phase (red/green) through V2I communication. To avoid boundary cases and ensure feasibility, the yellow phase is assumed to be a part of the green phase if the distance to the traffic light is close to the critical braking distance as formulated in \cite{gupta2019enhanced}. The time remaining in the current phase or transitioning to the next phase would change as the vehicle approaches the signalized intersection. In this section, $t_{\mathrm{GR},s}$ and $t_{\mathrm{RG},s}$ will represent time for a green-red and red-green transition respectively at position $s$, as shown in the Fig. \ref{fig_Traffic_Light_Phase}. In this work it is assumed that the positions of all the traffic lights in the route are known a priori from a navigation system, and contained in the set $\mathcal{D}_{\mathrm{TL}}$.

Assume that the discretized state dynamics for the eco-driving problem has the following form: 
\begin{equation}
    x_{s+1}=f_s(x_s,u_s), \quad s=1,\cdots, N-1.
\end{equation}
where $x_s\in\mathcal{X}_s\subseteq\mathbb{R}^p$ and $u_s\in\mathcal{U}_s\subseteq\mathbb{R}^q$ are the state and control action, respectively. 

In this work, the state variables are the vehicle velocity $(v_s)$, battery SoC $(\xi_s)$ and travel time $(t_s)$. The control actions are the engine torque $(T_{\mathrm{eng},s})$ and BSG torque $(T_{\mathrm{bsg},s})$:
\begin{subequations}
\begin{gather}
x_s=[v_s,\xi_s,t_s]^\mathsf{T}\\
u_s=[T_{\mathrm{eng},s}, T_{\mathrm{bsg},s}]^\mathsf{T}
\end{gather}
\end{subequations}
The equations describing the state dynamics $f_s(x_s, u_s)$ are:
\begin{subequations}
     \begin{gather}
         v_{s+1}=\sqrt{v_s^2+2\Delta d \cdot \left(\frac{F_{\mathrm{tr},s}-F_{\mathrm{road},s}(v_s)}{M}\right)}\\
         \xi_{s+1}=\xi_s-\frac{\Delta d}{\bar{v}_s}\cdot \frac{\bar{I}_{\mathrm{bat},s}}{C_{\mathrm{nom}}}\\
         t_{s+1} = \begin{cases} 
         t_s + t_{\mathrm{RG},s} &, s \in \mathcal{D}_{\mathrm{TL}} \text{ and } \bar{v}_s=0 \\
         t_s + \frac{\Delta d}{\bar{v}_{s}} &, s \notin \mathcal{D}_{\mathrm{TL}}   
         \end{cases} \label{eq: time dynamics}
     \end{gather}
 \end{subequations}
where $F_{\mathrm{tr},s}$ is the tractive force produced by the powertrain \cite{olin2019reducing}. $F_{\mathrm{road},s}$ is the road load resistive force, $M$ is the total vehicle mass, $\bar{I}_{\mathrm{bat},s}$ is the average current evaluated over a distance step, $C_{\mathrm{nom}}$ is the nominal battery capacity, $t_s$ is the travel time at a position $s$. Intuitively, Eqn. \eqref{eq: time dynamics} shows the time is teleported to the end of the red phase when the vehicle stops at the traffic light. 
 
The eco-driving optimization problem is formulated as a receding horizon optimal control problem where the full route of $N$ steps is solved over a reduced horizon $N_H (<<N)$. 
At any given position $s = 1, \dots, N-N_H$, the optimization problem is formulated as:
\begin{equation}
\begin{gathered}
\mathcal{J}^*(x_s)= \min_{\left\lbrace \mu_k \right\rbrace_{k=s}^{s+N_H-1}}\sum_{k=s}^{s+N_{H}-1} c(x_k,\mu_{k}(x_{k})) + c_\mathrm{T}(x_{s+N_H}), \\
c(x_k,\mu_{k}(x_{k}))=\left(\gamma \cdot \dot{m}_{\mathrm{f},k}(x_k,\mu_{k}(x_{k})) +(1-\gamma)\right)\cdot \Delta t_k
\end{gathered}
\end{equation}
where $\mu_k:\mathcal{X}\rightarrow\mathcal{U}$ is the admissible control policy at the step $k$ in the prediction horizon, $c:\mathcal{X}\times\mathcal{U}\rightarrow\mathbb{R}$ is the stage cost function defined as the weighted average of the fuel consumption and the travel time, $c_{\mathrm{T}}:\mathcal{X}\rightarrow\mathbb{R}$ is the terminal cost function. 

The state space and action space are subject to following constraints: $\forall s = 1,\dots,N-N_H$, $\forall k=s,\dots,s+N_H$:
\begin{subequations}
 \label{eq: constraints_N_H_horizon}
 \begin{align}
 v_k&\in[v_k^{\mathrm{min}},v_k^{\mathrm{max}}],\\
 \xi_k&\in[\xi_k^{\mathrm{min}},\xi_k^{\mathrm{max}}],\\
 t_k&\in\mathcal{T}_{\mathrm{G},k},\\
 a_k&\in[a^{\mathrm{min}},a^{\mathrm{max}}],\\
 T_{\mathrm{eng},k}&\in[T_{\mathrm{eng}}^{\mathrm{min}}(v_k),T_{\mathrm{eng}}^{\mathrm{max}}(v_k)],\\
 T_{\mathrm{bsg},k}&\in[T_{\mathrm{bsg}}^{\mathrm{min}}(v_k),T_{\mathrm{bsg}}^{\mathrm{max}}(v_k)],
 \end{align}
 \end{subequations}
 where $v^{\mathrm{min}}_s, v^{\mathrm{max}}_s$ are the minimum and maximum route speed limits respectively, $\xi_s^{\mathrm{min}}, \xi_s^{\mathrm{max}}$ are the static limits applied on battery SoC, $\mathcal{T}_{\mathrm{G},s}$ represents the feasible set on the travel time at the signalized intersection, $a^{\mathrm{min}}, a^{\mathrm{max}}$ represent the limits imposed on the acceleration for comfort, $T_{\mathrm{eng}}^{\mathrm{min}}(v_s),T_{\mathrm{eng}}^{\mathrm{max}}(v_s)$ are the minimum and maximum engine torque limits, and $T_{\mathrm{bsg}}^{\mathrm{min}}(v_s),T_{\mathrm{bsg}}^{\mathrm{max}}(v_s)$ are the minimum and maximum BSG torque limits, respectively. To ensure SoC neutrality over the entire itinerary, a terminal constraint $\xi_1=\xi_N$ is applied to the battery. 
 

\begin{figure}[t!]
	\centering
	\vspace*{0.15cm}\includegraphics[width=\columnwidth]{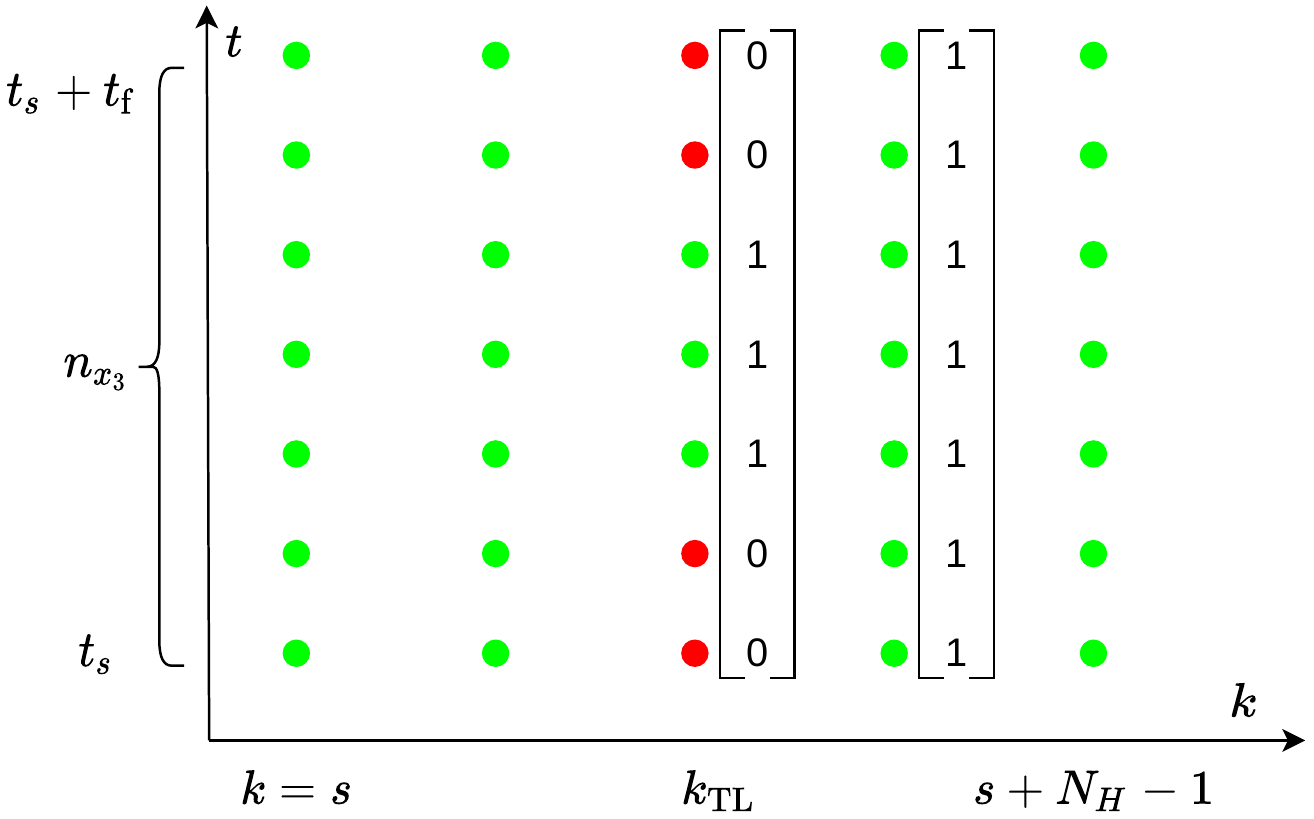}
	\caption{Travel Time Constraint at the Signalized Intersection in a $N_H$ Horizon.}
	\label{fig: Traffic_Light_Constraint_in_grid}
\end{figure}

 Consider a case where a $N_H$ horizon $(k=s,\dots,s+N_H-1)$ comprises of a signalized intersection located at $k_{\mathrm{TL}}$ as shown in Fig. \ref{fig: Traffic_Light_Constraint_in_grid}. The travel time at the beginning of the horizon is given by $t_{s}$ and a maximum travel time $t_\mathrm{f}$ is imposed at the end of the horizon. $t_\mathrm{f}$ is chosen such that the ego vehicle does not impede the traffic flow. This can be computed as an expectation from multiple historical trips along the same route \cite{jiang2013travel}. In this work, $t_\mathrm{f}$ is assumed to be $80$s for a 200m horizon to ensure feasibility. The status of the traffic light is then sampled into an indicator vector $\mathbb{I}_{\mathrm{G},k}$ of size $t_\mathrm{f}/\Delta t$ with each element representing no traffic light or green phase with 1 and red phase with 0. Note that $\Delta t$ here is the constant time discretization for sampling. The feasible set $\mathcal{T}_{\mathrm{G},k}$ is then defined as follows:
 \begin{equation}
     \mathcal{T}_{\mathrm{G},k} =
     \left\lbrace t:t_s+ \Delta t\cdot z | \left(\mathbb{I}_{\mathrm{G},k}\right)_z =1, z = 1, ..., t_{\mathrm{f}}/\Delta t \right\rbrace,
 \end{equation}
where $\left(\mathbb{I}_{\mathrm{G},k}\right)_z$ represents the $z^{th}$ element of the vector. The size of the sampled traffic light status vector is chosen to be the same as the grid size of the DP solver in the time dimension, which will be explained in the next section.

Knowing the driving conditions within the receding horizon, the optimization problem can be solved offline via DP. Following the nomenclature in \cite{sundstrom2009generic}, the optimal policy $\mu_k^*$, along with the optimal cost-to-go function $\mathcal{J}_k:\mathcal{X} \rightarrow \mathbb{R}$, $\forall k= s,\dots,s+N-1$ can be calculated through backward recursion as follows:
\begin{subequations}
    \begin{gather}
        \mathcal{J}_{s+N}(x) = c_{\mathrm{T}}(x) + \phi_\mathrm{T}(x), \\
    \mathcal{F}_k(x,u) = c_k(x,u) + \phi_k(x) + \mathcal{J}_{k+1}(f_k(x,u)),\\
    \mu_k^*=\argmin_{\mu_k} \mathcal{F}_k(x, \mu_k(x)),\label{eq: dp optimal policy}\\
    \mathcal{J}_{k}(x) = \mathcal{F}_k(x, \mu^*_k(x)),
    \end{gather}
\end{subequations}
where $c_k$ and $c_{\mathrm{T}}$ are the discretized stage and terminal cost function respectively; $\phi_k$ and $\phi_{\mathrm{T}}$ are penalty functions introduced to ensure that the trajectory stays feasible. In this paper, an approximate offline solution of a full-route optimization under partial information is initially obtained via DP and used as a terminal cost approximation. Other techniques, such as manual calibration or reinforcement learning techniques \cite{sutton2018reinforcement, zhu2020energy} may also be employed.
The benefit of using this method is that it provides the closed-loop optimal policy that inherently adds robustness against approximated plant dynamics or other modeling errors \cite{han2019fundamentals}. Further, DP can solve highly nonlinear and hybrid problems while ensuring constraint satisfaction.

\section{Architectures for DP Implementation} \label{sec: dp architectures}
\begin{figure}[!t]
    \centering
    \vspace*{0.15cm}\includegraphics[width=1\columnwidth]{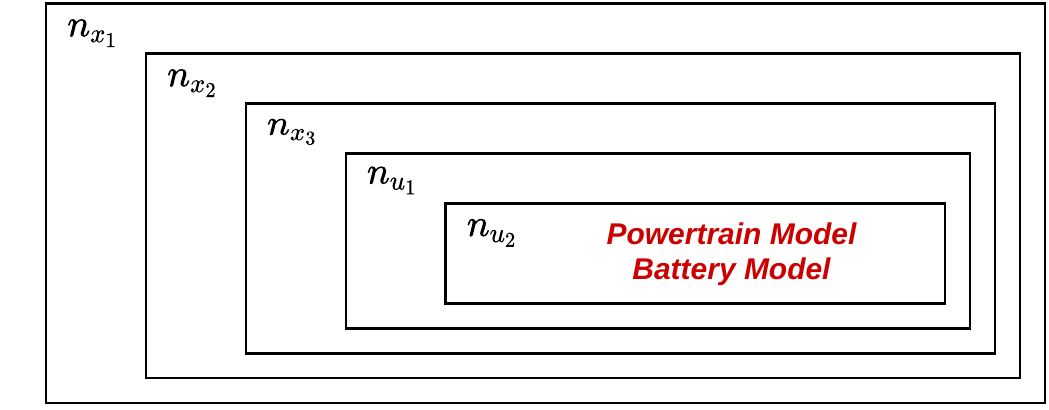}
    \caption{Serial DP Architecture}
    \label{fig:serial DP}
\end{figure}

For the eco-driving optimization problem, a serial DP architecture can be constructed, where each state-action combination is evaluated in a 5-layer nested loop, as shown in Fig. \ref{fig:serial DP}. 
The powertrain model takes the current vehicle speed, the engine torque and the BSG torque as the inputs and calculates the change in vehicle speed $\Delta v_{s}$, the change in time $\Delta t_s$, BSG power demand $P_{\mathrm{bat}}$ (input to the battery model) and the stage cost $c(x_s,\mu(x_s))$. 
The battery model takes $P_{\mathrm{bat}}$ and calculates the change in battery SoC $\Delta \xi$ for the nodes in the SoC grid. If an infeasibility is encountered for a state-action combination, the cost-to-go is set to a value that is orders of magnitude larger than the stage cost. 

To maintain a reasonable accuracy, the grid size is set to be $\left(n_{x_1},n_{x_2},n_{x_3},n_{u_1},n_{u_2}\right) = \left(35, 26, 40, 23, 30\right)$. As a result, the number of execution of the innermost loop is on the order of $10^7$ for each DP backward recursion step. Meanwhile, with $\Delta d=10m$, 20 recursion steps are required per distance step. Without any parallelization, the computational requirement becomes intractable for most automotive applications. 

\begin{figure}[!t]
    \centering
    \vspace*{0.15cm}\includegraphics[width=1\columnwidth]{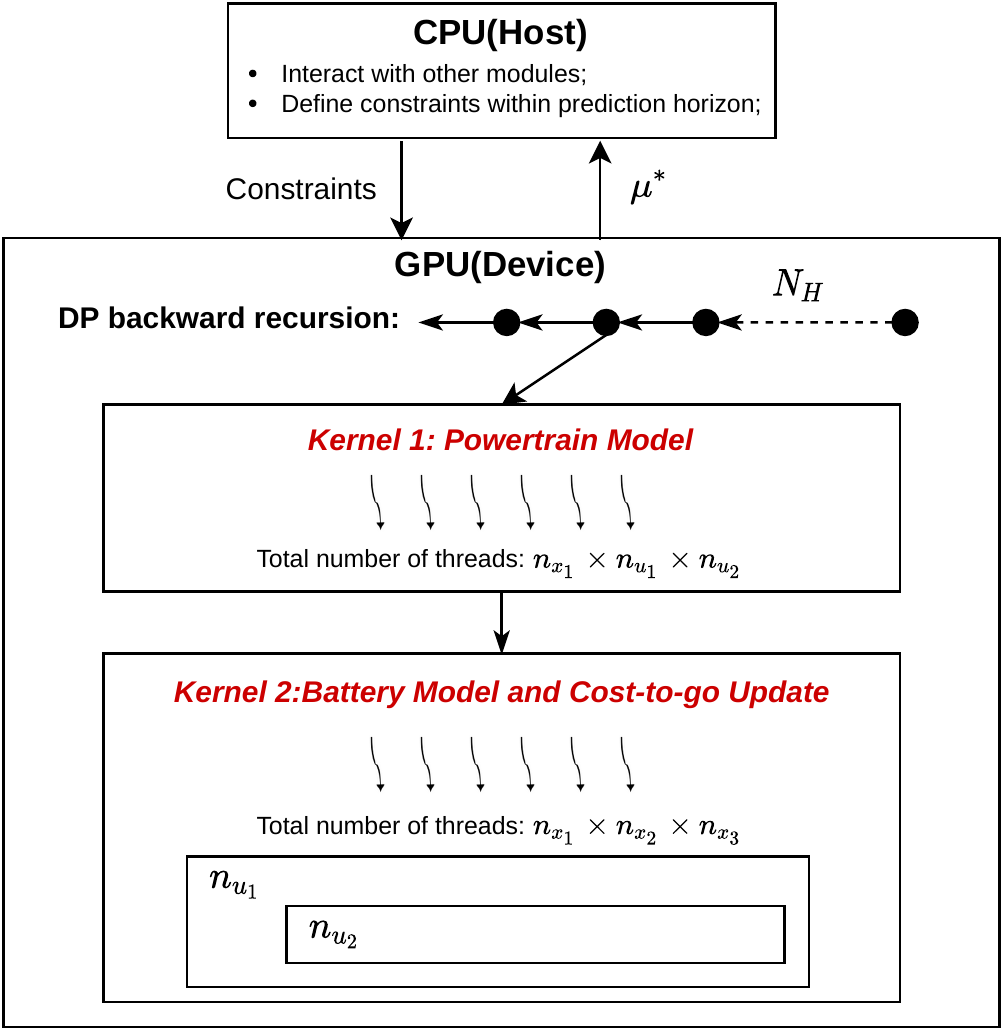}
    \caption{Parallel DP Architecture}
    \label{fig:parallel DP}
\end{figure}

To mitigate the issue, a parallel architecture is proposed in this work, and implemented on a NVIDIA GPU with CUDA C++ programming language. Fig. \ref{fig:parallel DP} shows the architecture of the parallel implementation. With comparable expense and power consumption, a CPU is faster at performing complex and long computing sequences thanks to the more resources dedicated to memory caching and control flow, whereas a GPU saves more resources on processors that can be run in parallel \cite{CUDA:2013}. 

In CUDA programming, CPU and GPU are referred to as host and device, respectively, while the Kernel is the function that is launched by the host and executed in parallel by the device. Each thread represents a sub-process running on GPU with unique inputs. To have an efficient implementation, the data transfer frequency between host and device should be minimized and, in the current framework, such data transfer occurs only once per DP optimization call. 

In the first Kernel, each thread represents a combination of $(v_{s}, T_{\mathrm{eng},s}, T_{\mathrm{bsg},s})$ on the 3D grid and calculates the corresponding $\Delta v_{s}$, $\Delta t_s$, $P_{\mathrm{bat},s}$ and $c(x_s,u_s)$ for the second Kernel. Each thread launched by the second Kernel represents a combination of the three states $(v_{s}, \xi_s, t_s)$, and it loops through the control inputs, determines $\Delta \xi$ based on $P_{\mathrm{bsg}}$ from the first Kernel, and finally determines whether to update the cost-to-go matrix and the optimal control matrix following \cite{sundstrom2009generic}. As Eqn. \eqref{eq: dp optimal policy} requires minimization to find the optimal control action, the nested loops for control actions are kept to avoid race conditions.

\section{Simulation and Evaluation of Results}
\begin{figure}[!t]
    \centering
    \vspace*{0.15cm}\includegraphics[width=1\columnwidth]{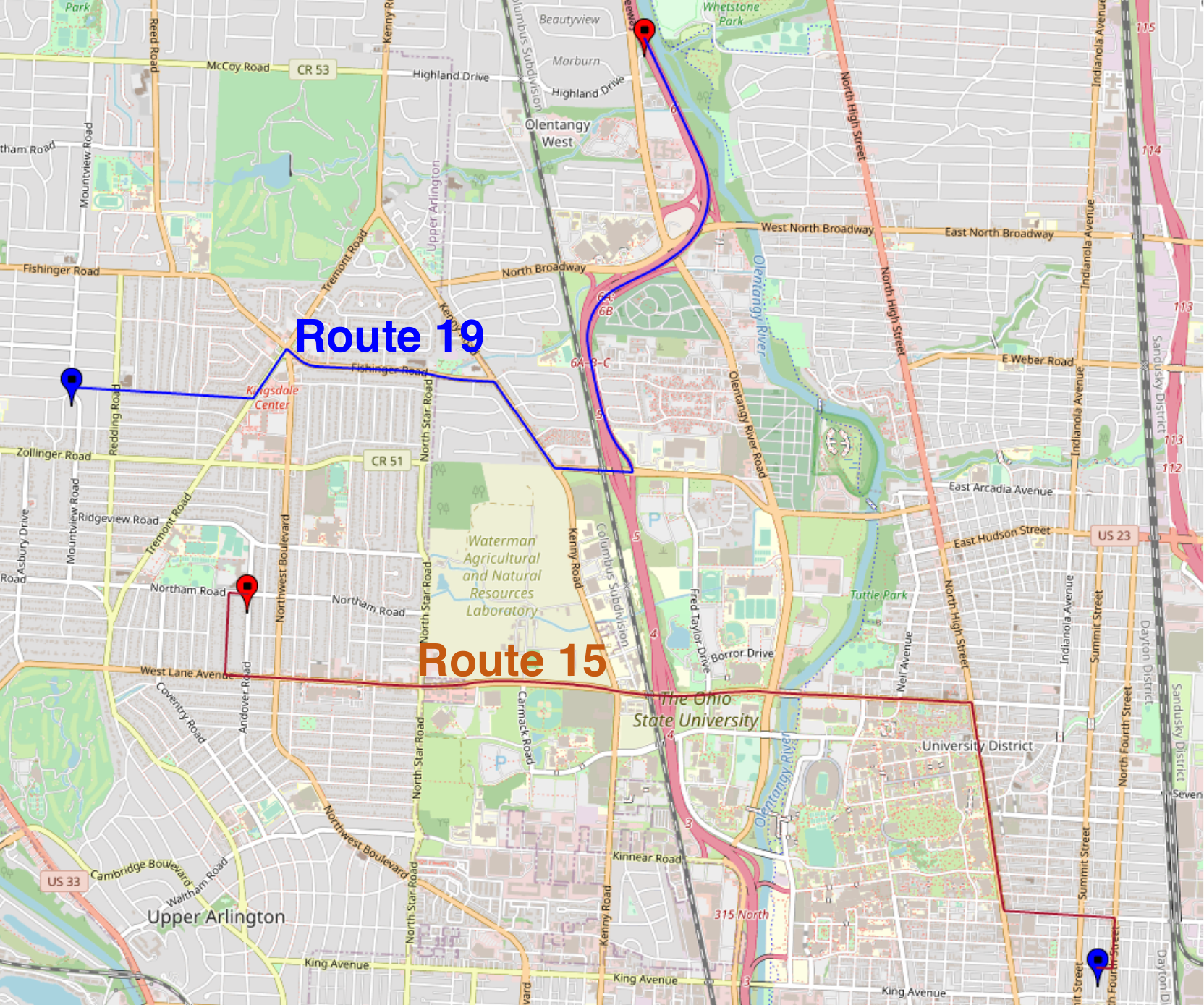}
    \caption{Route 15 and 19 on OpenStreetMap}
    \label{fig:Routes_on_OSM}
\end{figure}

In this paper, two real-world routes representing urban (Route 15) and mixed-urban (Route 19) driving conditions have been selected for the simulation and analysis. Both routes are located in Columbus, OH, USA as shown in Fig. \ref{fig:Routes_on_OSM} \cite{olin2019reducing}. 
The mixed-urban route is 7km in length and comprises 5 traffic lights and 3 stop signs, while the urban route is 7.5km in length and includes 22 traffic lights and 3 stop signs. 

For the comparison of results, a baseline controller was considered for the VD\&PT model. This controller consists of a heuristic Enhanced Driver Model (EDM) \cite{gupta2019enhanced} calibrated to control the vehicle velocity in such a way that the ego vehicle is able to follow a lead vehicle (when detected within a fixed line of sight) and to stop at traffic lights. The EDM parameters can represent different driver styles and can be tuned to match a real-world driver \cite{gupta2020estimation}. The production (rule-based) mHEV energy management controller was also considered in the VD\&PT model for the baseline case. 

\begin{figure}[!t]
    \centering
    \vspace*{0.15cm}\includegraphics[width=1\columnwidth]{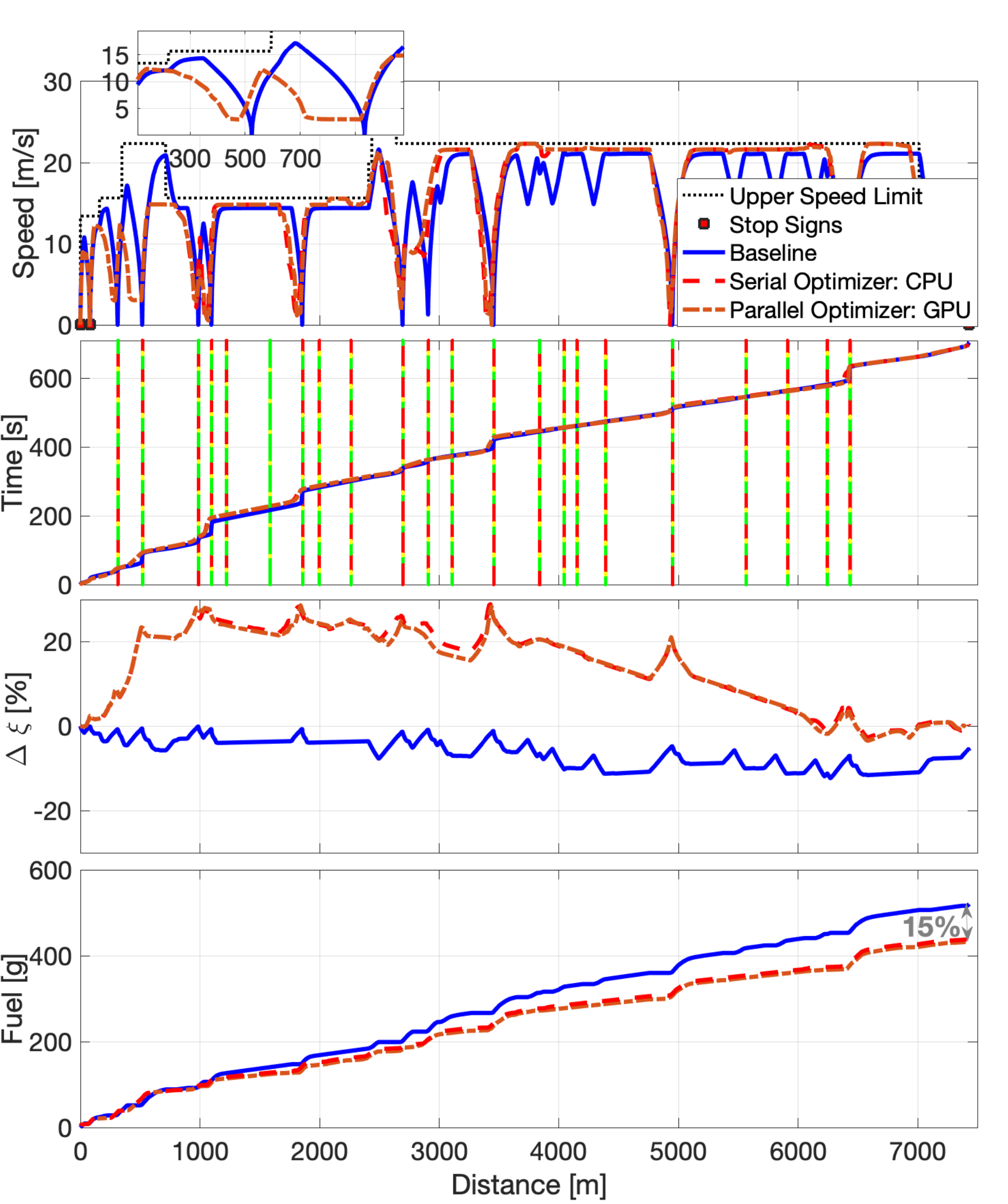}
    \caption{Comparison of VD\&PT State Trajectories and Cumulative Fuel Consumption on Urban Driving (Route 15)} 
    \label{fig:results urban trajectories}
\end{figure}

\begin{figure}[!t]
    \centering
    \vspace*{0.15cm}\includegraphics[width=1\columnwidth]{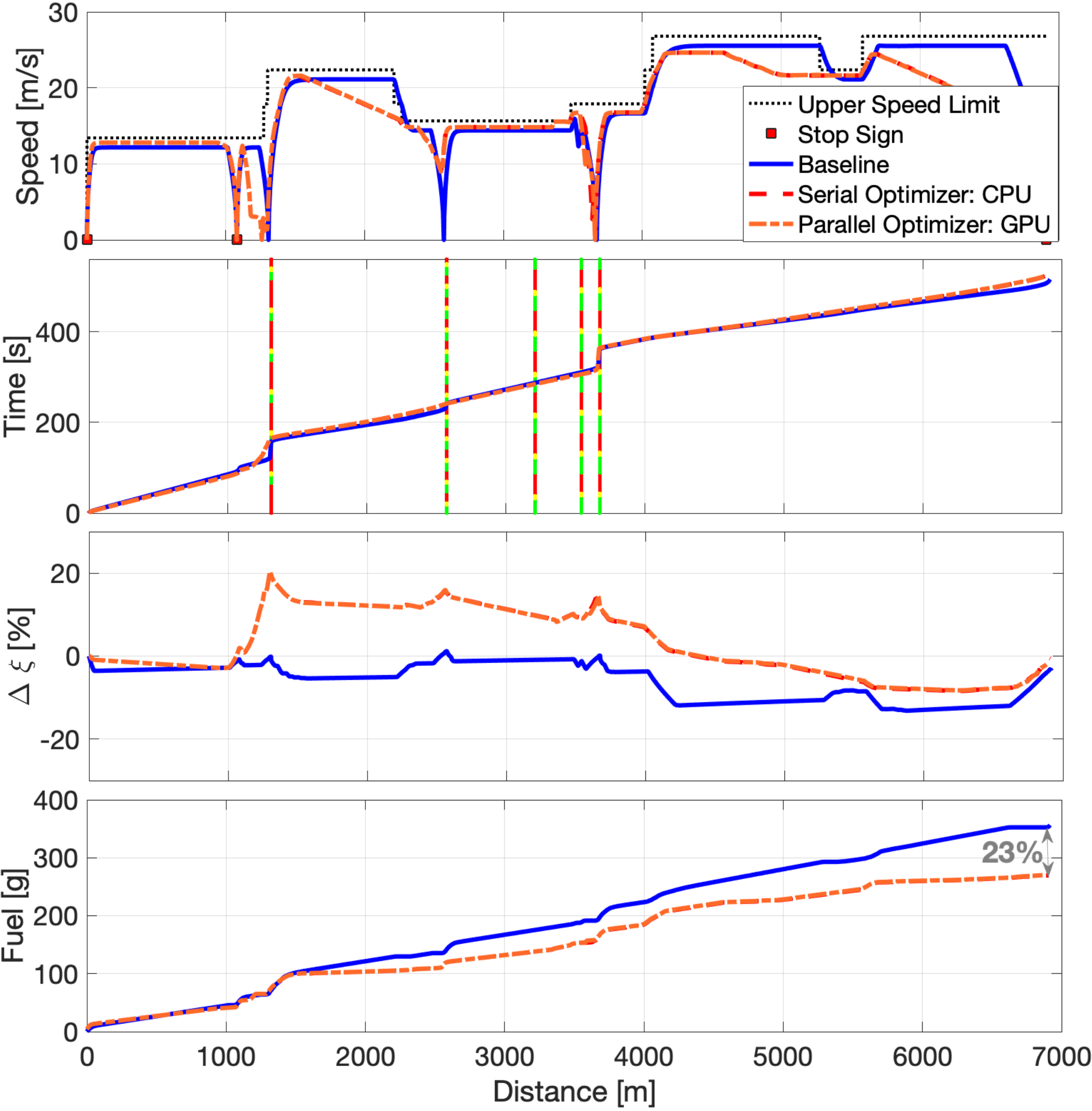}
    \caption{Comparison of VD\&PT State Trajectories and Cumulative Fuel Consumption on Mixed Driving (Route 19)}
    \label{fig:results highway trajectories}
\end{figure}

The 3-state DP-based MPC was then integrated into the vehicle simulator. As indicated above, a terminal cost approximation was obtained by computing a full route optimization offline via DP and storing the value function \cite{olin2019reducing}. Fig. \ref{fig:results urban trajectories} and Fig. \ref{fig:results highway trajectories} show the comparisons of the trajectories from the baseline and DP-based MPC controller on urban and highway driving, respectively. With comparable travel time, the optimized VD\&PT controller is able to reduce the fuel consumption by 15\% and 23\%, respectively, by optimizing the energy stored in battery and planning the speed trajectory in a way that avoids unnecessary braking events. Note that the serial and parallel implementation are also compared in Fig. \ref{fig:results urban trajectories} and Fig. \ref{fig:results highway trajectories}. As expected, no differences in the results are present in the two different implementations.

Table \ref{tab: solver time} shows the mean, variance and the maximum values of the solver time obtained by the serial and the parallel implementations. To perform a comparison of the computation throughput, both implementations were compiled in C++ code. The serial and the parallel implementations were then executed on a desktop PC with 2.9 GHz Intel Core i7 CPU and a NVIDIA RTX 2080 GPU, respectively. Besides the more than 90\% reduction in averaged solver time, the parallel implementation has a lower variance and maximum of the solver time as well. This is because the serial implementation has multiple break conditions in the nested loop and these conditions are dependent on the specific driving scenario and SPaT sequence encountered during the route. In the meantime, the parallel implementation goes through all the possible combinations by parallel threads, and thus is less subject to the driving conditions. Although both the CPU and GPU implementations benefit of the improved computational capabilities than typical rapid prototyping units, the benefits of utilizing GPU-based parallel computing are very appealing, especially considering that onboard GPUs are an indispensable development tool for the control of self-driving features in autonomous vehicles \cite{perez2020benchmarking}.   

\begin{table}[t!]
\caption{Solver Time Comparison}
\label{tab: solver time}
\centering
\begin{tabular}{|c|c|c|c|c|}
\hline
\multirow{2}{*}{} & \multicolumn{2}{c|}{Route 15} & \multicolumn{2}{c|}{Route 19} \\ \cline{2-5} 
                  & Serial       & Parallel       & Serial       & Parallel       \\ \hline
Mean (ms)             &           1600   &       98         &        1638      &      101          \\ \hline
Variance $(\text{ms}^2)$ &      112        &    11            &     287         &      13          \\ \hline
Maximum (ms)          &    2210          &      118          &         2312     &         123       \\ \hline
\end{tabular}
\end{table}

\section{Conclusion}
In this work, an eco-driving optimization problem was formulated for a mHEV with V2I communication and longitudinal automation. The problem was cast as a receding horizon optimal control problem with three states, namely vehicle velocity, battery SoC and travel time. A MPC implementation was then developed and solved using DP. Compared to previous studies where the predictive optimization is applied to control the vehicle velocity or to perform energy management of the hybrid powertrain, the work described in this paper integrates all elements of the eco-driving problem into a single controller. 

To accommodate the increased computational requirement from inclusion of additional states, a parallel implementation of DP is developed with CUDA programming on a NVIDIA GPU. The implementation reduces the computation time by more than 90\% (average, variance, and maximum of the solver time), compared to the serial counterpart. Simulation results indicate that the proposed controller decreases the fuel consumption by more than 15\% compared to the baseline controller, while keeping comparable travel time. 

Future work includes the implementation of the algorithm on rapid-prototyping systems and the study on how to optimally obtain the terminal cost function.

\section*{Acknowledgment}
The authors acknowledge the support from the United States Department of Energy, Advanced Research Projects Agency – Energy (ARPA-E) NEXTCAR project (Award Number DE-AR0000794).

\bibliographystyle{IEEEtran}
\bibliography{reference.bib} 
\end{document}